# Muon ID- Taking Care of Lower Momenta Muons

C. Milsténe, G. Fisk, A. Para
*FNAL, Batavia, IL 60510, USA*

In the Muon package under study, the tracks are extrapolated using an algorithm which accounts for the magnetic field and the ionization (dE/dx). We improved the calculation of the field dependent term to increase the muon detection efficiency at lower momenta using a Runge-Kutta method. The muon identification and hadron separation in b-bbar jets is reported with the improved software. In the same framework, the utilization of the Kalman filter is introduced. The principle of the Kalman filter is described in some detail with the propagation matrix, with the Runge-Kutta term included, and the effect on low momenta single muons particles is described

## 1. INTRODUCTION

Muons in the 3-5 GeV range are a substantial part of the μ produced at 500 GeV center of mass energy, for example they constitute ~14% of the muons produced in bbar-b events. This note describes improvements in the μ identification at the lower momentum limit. The improvements are made by applying two techniques: first, second order corrections to the propagation algorithm are introduced using the Runge-Kutta method, second, using a Kalman filter, multiple scattering and random processes are added to the propagation. Multiple scattering is particularly important for low momenta tracks, hence of particular interest in the region explored, The note describes the improvements achieved by these methods

## 2. THE RUNGE-KUTTA CORRECTION

The Muon package has been described in detail Ref.[1]. Also, the Muon ID algorithm has been applied to 10000 b-bbar Events [2]. This note focuses mostly on the simulated muons of lower momenta which still reach the muon detector (3GeV≤ p <5GeV) and for which the approximation $\Delta p_T / \Delta t \sim dp_T / dt$, is insufficient. Here $\Delta p_T$ (GeV/c) is the variation in the transverse momentum of a particle going through a magnetic field B (Tesla) for a time $\Delta t$(s) and

$$dp/dt = 0.3q \, \bar{v} \times \bar{B} = \alpha \, \bar{p} \times \bar{B} \, , \text{ with } \alpha = 0.3(1/E)qc_{light}$$

E is the particle energy in GeV, q its charge in electron units and $c_{light}$ (m/s). In a 5 Tesla magnetic field, for low momenta, one has to calculate the integral in order to obtain the finite difference equation of motion:

$$\Delta p_T = \int \alpha \, \bar{p}_T \times \bar{B}_z \, dt$$
$$\Delta p_z = 0.$$

and the magnetic field dependent change in momentum is now given by the equations

$$\Delta p_T = (\alpha / \delta)(\bar{p}_T \times \bar{B}_z) - \eta p_T$$
$$\delta = 1 + 0.25\alpha^2 B_z^2 \Delta t; \; \eta = 0.5\alpha^2 B_z^2 \Delta t / \delta$$

This introduces correction terms in the magnetic field dependent momentum change.

The term - η · pi ; (i=x ,y) is a friction term. The energy loss due to ionization in matter (for step n of the trajectory) is given by $\Delta p_{i\,Matter} = (dE(n)/dx).(E(n)/p).(p_i(n)/p(n)).ds$, $i = x,y,z$

Here ds is the distance traversed by the particle. 10000 bbar-b jets events were analyzed with the improved software.

## 3. MUON-IDENTIFICATION WITH THE IMPROVED METHOD

### 3.1. The Algorithm

The analysis starts with reconstructed and well fitted tracks from the tracker with energy deposit in the Electro-Magnetic Calorimeter, EMCal, the Hadron Calorimeter, HDCal and MUDet within a (θ,φ) angular width around their path. The angular width is optimized for each sub-detector [1],[2]. The Muon Detector trajectories are selected if #hits ≥12 in 12 or more layers within the (θ,φ) width. For tracks below 10 GeV a (1/p) dependent bin has been chosen. Also, tracks which do not leave a signal in the last 5 layers of HDCal are rejected. This last cut takes advantage of the fact that most of the hadrons interact well before the end of HDCal and that the hadron showers are short whereas the muons have a penetrating power. Therefore the muons, reach the end of HDCal and leave hits in the 5 last layers. An additional cut requires the layers with the minimum number of hits to have 1 or 2 hits, another characteristic of minimum ionizing particles(MIP). This scans the track for regions of low activity in which MIP pattern can be checked.

### 3.2. Performance of Muon ID

. The effect of the cuts is shown in Table I. The detection Efficiency is actually better than reported because some of the undetected tracks share hits between barrel and end-cap and will be detected in a 4π detector. Also, the normalization is to the total number of reconstructed-final-fitted tracks, whereas part of these do not reach 12 layers (see above) and should be subtracted.

Table I: Muon Identification in 10000 b-bbar with at least 12 layers/12 Hits in the Muon Barrel

| Detector | Muons | Pions | Kaons | Protons |
|---|---|---|---|---|
| Tracker Recons. Final | 739 | 18024 | 4303 | 1712 |
| Tracker Good Fit | 715 | 17120 | 4072 | 1579 |
| HDCal    1 ≤ min Hits ≤ 2 | 700 | 588 | 247 | 26 |
|           5 –Last Layers > 0 Hit | 700 | 357 | 204 | 15 |
| MUDet, ≥ 12 hits ≥ 12 layers | 671 | 77 | 50 | 5 |
| MUDet -  Min Hits ≤ 2 ; Max Hits ≤ 7 | 670 | 59 | 39 | 5 |
| Efficiencies of µ Detection And  Hadrons Rejection | Det. 94% | Rej. 1/305 | Rej. 1/110 | Rej. 1/342 |

The overall muon detection efficiency obtained without discounting for the effects described above is ~94%. This covers the entire momentum range. It is ~100% at, and above, 10 GeV. For muon momentum at or above 3GeV the detection efficiency increased with the Runge-Kutta correction from ~30% to 66%. A Kalman filter is used next in order to account for multiple scattering and other random processes and get a better separation of muons from hadrons.

### 3.3.  The Principle

The Kalman Filter is a method used to construct processes for which the information at each step can be fully derived from the information at the preceding step and where a covariant matrix of error is available at each step. The Kalman Filter is composed of two components. The propagation component, using the propagation matrix, propagates the information and the error information at each step. The filtering component combines the information received at the end of the step to the information from data measurements.

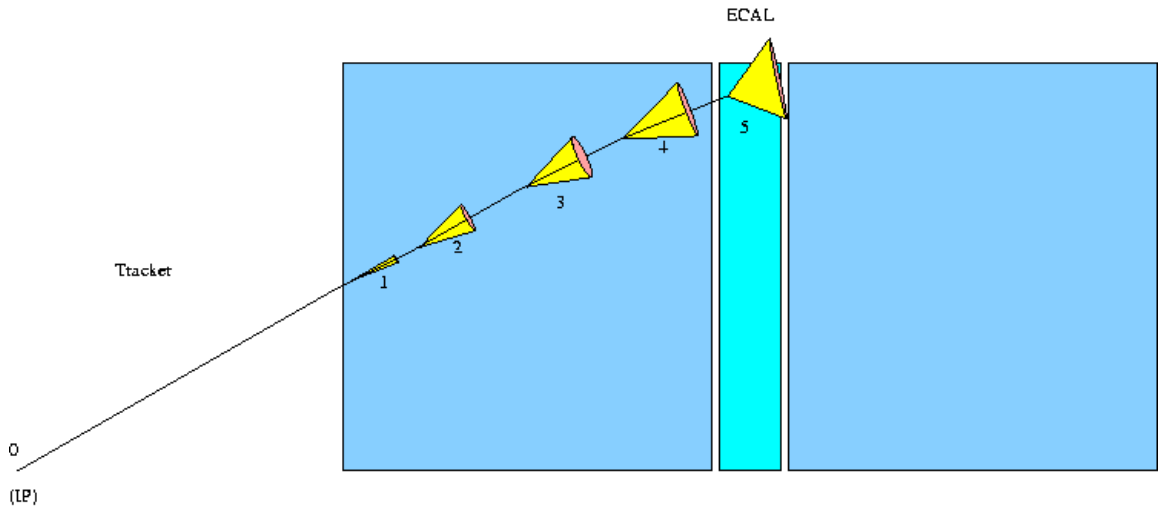

Figure 1: Stepping in the Passive Material (1, 2, 3,4, ) and recording in the Active Material step 5

The Kalman Filter operates by using a "State Vector" that describes the processes' history. For the present case, a phase space point is used as the State Vector. The State Vector change includes the effects of dE/dx and of the magnetic field in a step by step procedure. It is calculated using the transport matrix (steps 1, 2, 3, 4 shown in Figure 1) in the passive material. The same analytical form is used in the stepper, but is now translated into a transport matrix, with the Runge-Kutta term included. The state vector at location k-1 is propagated using the propagation matrix, together with the covariant matrix, represented by the cone, which provides the information on random processes,. Following the propagation step, the State Vector is updated using Kalman filtering. The filtering combines the information derived from the propagation with the measurement made at the new point to produce an optimal state vector. It takes place when the hit is recorded in the <u>active material</u> (step 5) as shown in figure 1. The procedure defines, for each radius, a ($\varphi$, $\theta$) dynamic path in which the hits left by the particle in the active material of the sub-detectors are the data collected.

### 3.4.  The Equations

The stages of the Kalman Filter are described by the equations below.

$$\begin{cases} \vec{x}_k(-) = \Phi_{k-1} \cdot \vec{x}_{k-1}(-) \\ P_k(-) = \Phi_{k-1} \cdot P_{k-1}(-) \cdot \Phi_{k-1}^T + Q_{k-1} \end{cases}$$ (1) Propagation in passive materiel

$$\begin{cases} \vec{x}_k(+) = \vec{x}_k(-) + K_k \cdot [\vec{z}_k - H_k \cdot \vec{x}_k(-)] \\ P_k(+) = [1 - K_k \cdot H_k] \cdot P_k(-) \end{cases}$$ (2) K.Filter, in scintillator,

where the signal is recorded

$$K_k = P_k(-) \cdot H_k^T \cdot [H_k \cdot P_k(-) \cdot H_k^T + R_k]^{-1}$$ (3) K.Gain Matrix

$$Q_{k-1} = |\vec{p}| \cdot \Theta_0 \cdot I$$

Qk is the noise from Multiple scattering- a [6x6] matrix, Rk is the measurement error, dx, dy, dz - a [3x3] matrix

Hk is the measurement matrix - a [3x6] matrix, Φk = propagation matrix, applied in passive material- [6x6] The matrix, Xk(-) is the extrapolated vector state (x, y, z, px, py, pz) -[6x1] , Zk is the measured quantities (Φ,θ,r) translated into (x,y,z)-[3x1] matrix, Xk(+) is the state vector after applying the Kalman filter-[6x1] Matrix,Kk is the Kalman Gain matrix - a [6x3] Matrix

The transport matrix is given below in some detail. The matrix represent the propagation equations given in [1] and [2], dT is the time taken by step n., I is a [6x6] matrix Unity. The term - η · pi ; i=x ,y is a friction term is in the diagonal in the Transport Matrix

$$\begin{Bmatrix} x_k(-) \\ y_k(-) \\ z_k(-) \\ px_k(-) \\ py_k(-) \\ pz_k(-) \end{Bmatrix} = \left( dT * \begin{pmatrix} aa & ab \\ ba & bb \end{pmatrix} + I \right) * \begin{Bmatrix} x_{k-1}(+/-) \\ y_{k-1}(+/-) \\ z_{k-1}(+/-) \\ px_{k-1}(+/-) \\ py_{k-1}(+/-) \\ pz_{k-1}(+/-) \end{Bmatrix} \quad ; \quad \phi = dT * \begin{pmatrix} aa & ab \\ ba & bb \end{pmatrix} + I$$

$$ab = \begin{pmatrix} (cdedx) & f(Bz) & 0 \\ -f(Bz) & (cdedx) & 0 \\ 0 & 0 & (cdedx) \end{pmatrix} \quad ; \quad bb = \begin{pmatrix} dxyz & 0 & 0 \\ 0 & dxyz & 0 \\ 0 & 0 & dxyz \end{pmatrix} \quad ; \quad \text{cdedx} = dE/dx \cdot 100 * c/\text{Pabs}$$

$$R_0 = \begin{pmatrix} dx & 0 & 0 \\ 0 & dy & 0 \\ 0 & 0 & dz \end{pmatrix} \quad ; \Theta_0 = (13.6 MeV/P \cdot \beta c)\sqrt{x/X0} \cdot (1 + 0.038 \cdot lan(x/X0)); \quad \beta = c = 1$$

$$x = r \cdot \sin\Phi; \; y = r \cdot \cos\Phi; \; z = r \cdot ctg\Theta$$

$$dxyz = 100 * clight/E_n; \quad f(Bz) = 0.3 * q * \frac{p_x(n)}{E(n)} * clight * B_z \quad distance \text{ in (cm), energy in GeV, B(Tesla)}$$

The state vector at location k-1, is propagated using the propagation matrix, to location k. The choice of the state vector as the phase space point allows the use the stepper Algorithm translated into the propagation matrix. In the notation adopted the (-) indicates the extrapolation and the (+) the Kalman weighting.

### 3.5. The Results

The results for single muons at 4GeV before and after the application of the Kalman Filter are shown in Figure 2.

The figure below shows $\Delta\varphi$ (track-hit) between the extrapolated track and the hit at different depths (different layer) in the Muon Detector. In the LHS of the figure the width of the distribution is driven by the resolution at 4 GeV and the hit collection path covers ~ 3bins in φ. The RHS of the figure shows that one can restrict the hit collection to one bin in φ. The Filter helps to master random effects such as multiple-scattering when particles lose sufficient amount of momentum and enters the domain where these effects are dominant. As a result, the Filter allows a better separation between muons and neighboring hadrons in jets.

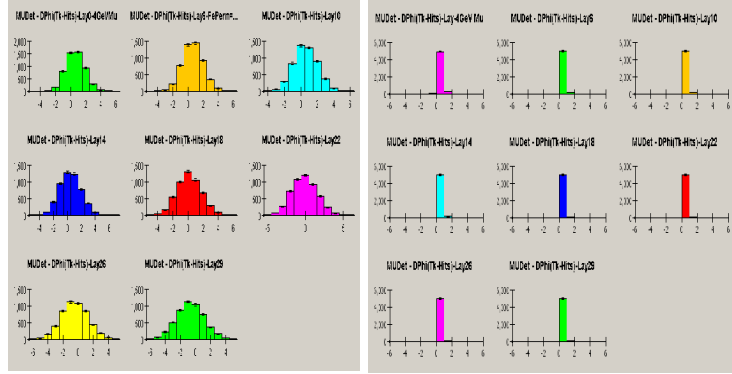

Figure 2: $\Delta\varphi$(track-hit) at 4 GeV Before and After applying Kalman Filter

## 4. CONCLUSIONS

The second order corrections and the Kalman filter were implemented in the Muon code to improve the efficiency in the low momentum range. The improved analysis includes multiple scattering, the magnetic field effect integrated using the Runge-Kutta method and the loss of energy through ionization, dE/dx. The error at starting point has been chosen to be the angle bin size in the calorimeter ECAL. Prior to the entry into ECAL the paths are constructed by the stepper. The Kalman filter reconstructs realistic propagation at each step and allows hits to be collected in a narrower kinematic band. As a result of applying the Kalman Filter, the cuts around the track path may be restricted to a minimum even for low momenta. This was previously possible only at higher momenta. This new capability is particularly important to allow for an effective separation of particles in jets.